# Superconductor Optical Devices Modified By Patterned Illumination

Hamed Kamrani, Mohammad Saeed Sharif Azadeh, Alireza Kokabi, and Mehdi Fardmanesh, *Senior Member, IEEE*

***Abstract*-We show that a thin superconductor slab illuminated by a desired pattern of light behaves as a completely controllable two-dimensional photonic media that could be applied in a wide range of optical devices. In this case, the permittivity spatial variation, which is fundamentally required in many photonic devices, is achieved by means of selective variation of cooper-pair density under patterned irradiation. The process of photo-effect in superconductors is the proposed mechanism for deformation of spatial distribution of cooper pair density and consequently non-uniform permittivity. In this perspective, the effects of non-uniform photon irradiation on the density of paired carriers and two-dimensional position dependent permittivity are investigated. Applying these results, the optical properties of several illumination patterns leading to the formation of different types of optical devices are studied. As we will show, in contrast to previous superconductor photonic crystals, the proposed structure has promising advantages such as possibility of implementing tunable heterostructures, optical gratings, and permittivity gradients with significant simplicity.**

I. INTRODUCTION

The optical properties of the superconductors have attracted a great deal of attention due to its potential applications in high-speed optoelectronic devices. Especially, incorporating the photonic crystals concept into superconductor electronics has opened novel approaches for tuning the photonic crystal (PC) properties such as transmission wavelength. [1-9] In this regard, Chen et al.[3] have investigated the control of photonic band gap (PBG) of the superconductor-dielectric lattice by variation of superconductor operation temperature. In the mentioned method, the PBG can be precisely controlled by the variation of the PC temperature. However, the inertia of temperature variation is not favorable for the applications such as optical switches that require rapid permittivity change. Furthermore, in the mentioned approaches, the background material is assumed to be non-superconducting, which may cause fabrication complexities. Actually, from the fabrication point of view, the best structures are the simplest ones. Therefore, it might be extremely interesting to apply an entire superconducting media as an optical device and induce the desired modifications nondestructively. Motivated by this fact, the Abrikosov lattice as a periodic structure of the permittivity has previously proposed to be a tunable PC whose properties can be adjusted by the applied magnetic field. In this method [5], penetration of magnetic field in the superconducting media suppresses the superconductivity inside the vortices and makes the modification of permittivity possible. However, the configuration of the Abrikosov lattice and its corresponding distribution of the permittivity are not fully controllable. Moreover, the variation of the magnetic field between neighboring vortices is smooth in the range of penetration length, which makes it difficult to form nanoscale PCs especially in the case of high $T_c$ superconductors. In contrast, since in high $T_c$ superconductors, coherence length is in the order of sub-nanometer, the boundary of the superconducting and non-superconducting zones is also very sharp. This promising characteristic of high $T_c$ superconductors makes them qualified for nanoscale PCs provided that a mechanism would be established to impose the superconductivity suppression without any physical contact or fabrication complexities.

While light patterning is a crucial part of the modern nanotechnology fabrication processes, it seems that it could also play the main role in the induction of the modified regions in the hosting superconducting media with ultra high spatial resolution. In this approach, the superconductivity suppression is achieved through the process of the cooper-pair breaking caused by photon absorption. Consequently, it is more feasible to adjust the optical properties of superconductor photonic crystal (SCPD) and other superconductor-based optical devices with higher accuracy. Thus, a wide range of optical devices can be synthesized on a uniform superconductor with the advantage of capability of switching overall device operation simply by changing the pattern of illumination. In the other words, there would be a device with the property of behaving as a PC or as an optical resonator for instance, by switching the incident radiation pattern. It is remarkable that the photon energy of the transmitted electromagnetic wave, unlike the patterning irradiation, should be less than superconductivity specific gap, 2Δ. In fact, the frequency of the transmitted electromagnetic wave should be in the range that the depairing does not occur. Simultaneously, the frequency of the patterning radiation should be sufficiently high enough to be able to diminish the superconductivity.

In contrast to the method described in [5] in which the obtained structure is uniform periodic ones, here, there would be the capability of implementing non-periodic structures such as light-guides, optical resonators and optical filters, which are also investigated in this work.

We also propose the application of the photon induced pair breaking process in the implementation of controllable optical devices such as tunable PCs.

Here, we applied this method for theoretical implementation of a wide range of optical devices. First, the T* model is applied for modeling the process of the photon-induced cooper-pair depairing. Then, introducing this model in the Drude permittivity of the superconductors, the optical characterization of the proposed devices is analyzed.

## II. Theoretical Model

Originally, the optical response of a superconducting material can be classified into either thermal or non-thermal. The former is through a variation in the biased state resistivity, which is the operation principle of the bolometric detectors [10]. Obviously, the thermal effect is dominant at T~Tc where the resistivity varies sharply with the temperature. On the other hand, the non-thermal effect is based on the dependence of the gap parameter to the density of the quasi-particles generated by photo-absorption. This effect is dominant at temperatures far below $T_c$ provided that the superconductor slab would be much thinner than the absorption coefficient so that the light intensity can be assumed uniform. Here, we use T* model [11] to estimate the density of quasiparticles generated by the illumination. This model is initially motivated by the rate equations of Rothwarf and Taylor [12]

$$\frac{dN}{dt} = I_0 + 2N_\omega/\tau_B - RN^2$$

and

$$\frac{dN_\omega}{dt} = RN^2/2 - N_\omega/\tau_B - (N_\omega - N_{\omega T})/\tau_\gamma \tag{1}$$

In which $N$ denotes the number density of quasiparticles, $I_o$ is the volume rate of quasiparticles creation by an external mechanism, $N_\omega$ is the number density of phonons with energy greater than $2\Delta$, $\tau_B^{-1}$ is the mean rate at which these phonons create quasiparticles, $R$ is a recombination coefficient, $\tau_\gamma^{-1}$ is the rate at which phonons of energy greater than $2\Delta$, disappear by processes other than quasiparticles creation, and $N_{\omega T}$ is the thermal equilibrium number density of phonons with energy greater than $2\Delta$. According to T* model, the Rothwarf -Taylor equations can be solved to obtain the steady-state solution as

$$\Delta N = N_T\left[\left(1 + \frac{I_0 \tau_{\text{eff}}}{N_T}\right)^{1/2} - 1\right] \tag{2}$$

where

$$\tau_{\text{eff}} = \tau_R\left(1 + \tau_\gamma/\tau_B\right) \tag{3}$$

In which $\Delta N$ is the excess number density of quasiparticles, $N_T$ is the thermal equilibrium number density of quasiparticles, and $\tau_R = (RN_T)^{-1}$ is the intrinsic recombination time.

In the case of light emission, with monochromatic radiation of frequency $v$, $I_0$ is obtained by

$$I_0 = (hv/E)(FP/u)(1/hv) \tag{4}$$

Where $E$ is the average energy of an excited quasiparticle, $hv/E$ is the average number of quasiparticles produced per photon if all of the absorbed optical energy is only shared by quasiparticles, $F$ is the fraction of the absorbed optical energy that is shared among the excess quasiparticles, $(hv)^{-1}$ is the number of photons per unit energy, $P$ is the absorbed optical power, and $u$ is the effective volume of the superconductor. It is assumed that the surface of the superconductor is uniformly illuminated and that the superconductor is sufficiently thin that the excess number of quasiparticles, $\Delta N$, is spatially uniform. After substitution of Eq. (5), with the phonon trapping limit, where $\tau_{\text{eff}} = N_T \tau_\gamma / 2N_{\omega T}$ equation (3) becomes

$$\Delta N = N_T\left[\left(1 + \frac{F \tau_{\text{eff}}}{N_T E u}P\right)^{1/2} - 1\right] \tag{5}$$

Equation (5) and a large number of experimental works [13,14] show that density of quasiparticles can be controlled by the intensity of the incident light.

On the other hand, effective complex permittivity of superconductors is well described by the two-fluid and Drude models through the following equation:

$$\varepsilon_{\text{eff}} = \varepsilon(x,y)\left[1 - \frac{\Omega_{sp}^2(x,y)}{\Omega^2} - \frac{\Omega_{np}^2(x,y)}{\Omega(\Omega + i\gamma(x,y))}\right] \tag{6}$$

In which $\gamma(x,y)$ is the damping term of the normally conducting electrons, $\varepsilon(x,y)$ is the superconductor relative permittivity, $\Omega_{sp}$ and $\Omega_{np}$ are plasma frequencies for the superconducting and normal carriers, respectively, given by

$$\Omega_{sp}(x,y) = \sqrt{\frac{[N_0 - N(x,y)]e^2}{m\varepsilon_0 \varepsilon}} \tag{7.a}$$

$$\Omega_{np}(x,y) = \sqrt{\frac{N(x,y)e^2}{m\varepsilon_0 \varepsilon}} \tag{7.b}$$

Where $N_0$ is the total number of electrons, $e$ is the electron charge and m is the electron mass. According to this equation, it is obvious that applying appropriate photon irradiation on the superconducting material, drastically changes the effective value of the complex permittivity. Thus, selective irradiation leads to spatial distribution of complex permittivity and consequently superconductor uniform media behaves as an electro-optically non-uniform media, which is ideal for making photonic crystal and its related optical devices.

## III. RESULTS AND ANALYSES

We investigate the effects of two-dimensional irradiation on the superconducting slab and implementing photonic device structures. Various types of optical devices which show different benefits of our method are proposed and simulated. In our simulation, we considered the values of $\varepsilon(x,y)=10$, $\gamma=10^{13}$ and $N_0=1.6\times10^{26}$ for complex permittivity constants and use FEM method for simulating the behavior of optical devices.

A) Implementing Photonic Crystal

First, we investigate the effect of irradiation on the two-dimensional photonic band structure of uniform superconducting slab. The typical considered structure, forming a rectangular 2D photonic crystal, is shown in Fig. 1.

The whole structure is homogeneous superconducting material in which the superconductivity of the shaded rods is partially suppressed by the illumination. The lightening pattern can also be vice versa, in a way that the shaded rods are masked and the remaining parts are illuminated which leads to superconductivity suppression in these areas. These two complementary illumination patterns result in the formation of two possible kinds of PCs called inner and outer respectively.

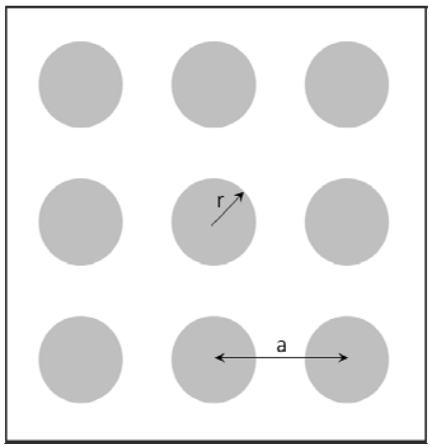

Fig. 1. Typical two-dimensional rectangular photonic crystal.

In Fig. 2, the TE-mode band structure of a rectangular photonic crystal for the inner configuration, with $r/a$=0.3 is presented. In this structure, the superconductivity of the rods is assumed to be fully suppressed due to the irradiation. The photonic band structure clarifies that irradiation leads to gap opening in the uniform superconducting media with the normalized gap width of $\Delta\omega$=0.108 and normalized gap center frequency of $\omega_c$=0.361.

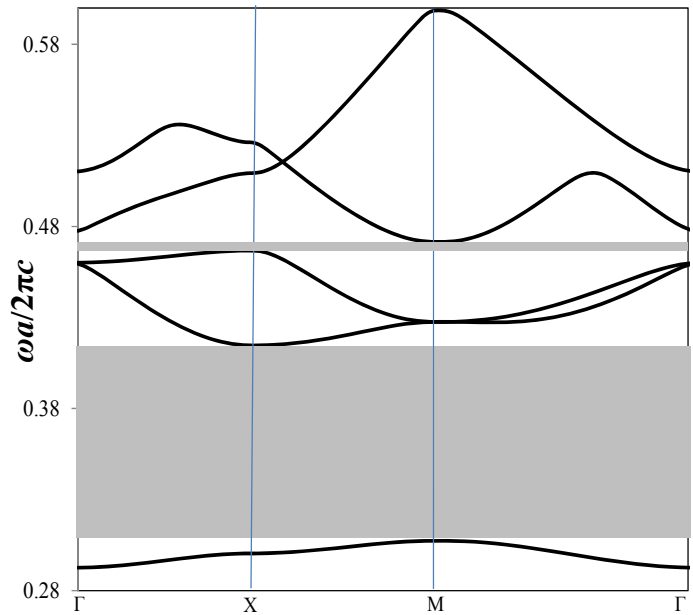

Fig. 2. TE-mode photonic band structure of a rectangular superconducting photonic crystal

Since the illumination pattern can be widely adjustable, by changing the radius of illuminated rods, here, we consider the effects of different fill factors, $r/a$, on the center and width of the photonic band gap. Fig. 3.a shows this dependency for the case of rods with fully suppressed superconductivity. The band gap of the photonic crystal could also be adjustable by changing the permittivity of rods using different illumination intensities. Fig. 3.b, showing the variation of photonic band gap with respect to normal electrons density, indicates that the photonic band gap is increased by increasing the intensity of illuminated light. This result was predictable since we expect the enhancement of permittivity contrast between the dark and bright regions by increase in the irradiation intensity.

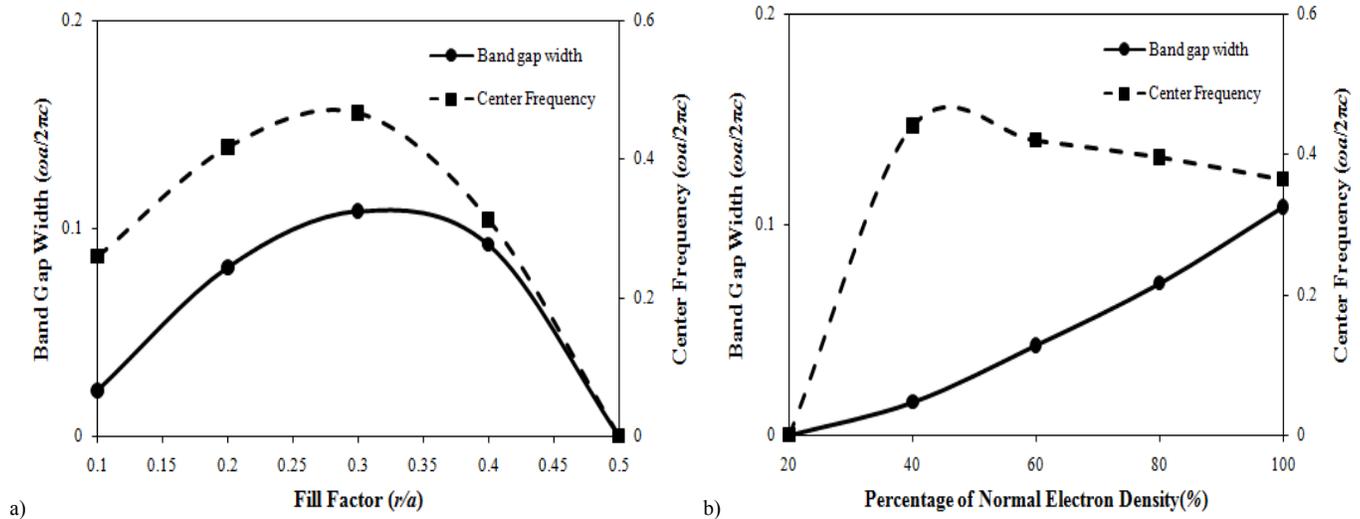

Fig. 3. Photonic band gap width and center frequency for different values of a) fill factors and b) normal electron densities

Another way to control the photonic band gap is the modification of the lattice structure. As mentioned before, unlike the other existing methods, we can change the photonic lattice by applying different illumination configurations. The photonic band gap for some different configurations is presented in Table .1. These results clarify the effect of lattice structure on the photonic band gap width and center.

TABLE I
Dependency of photonic band gap width and center frequency to lattice structure

|  | Rectangular lattice | Triangular lattice | Hexagonal lattice |
|---|---|---|---|
| Band gap width | 0.108 | 0.122 | 0.160 |
| Center frequency | 0.361 | 0.370 | 0.395 |

B) Implementing Photonic Devices

The major capability of the proposed method is the flexibility of the illumination pattern and easily applicability of its modification. The simplest way to switch from photonic crystals to the other types of photonic devices is imposing vacancies and interstitials in the photonic crystal structure. This approach is applied in Fig. 4.a which presents a manipulated square lattice containing special structure of ring resonator. [15] In this structure, designed for the 1550nm communication window, the radius of the rods is assumed to be 100 nm with the spacing of 540 nm along with the superconductivity inside the rods is assumed to be fully suppressed. Fig. 4.b shows the propagation of a special wavelength applied to port A of the ring resonator which is completely transmitted to the port D. The path of the transmitted electromagnetic wave is illustrated by its polarization graph of the electric field. As it is obvious, very high spectral selectivity can be achieved by this high quality factor resonator and the intrinsic single mode nature of the ring.

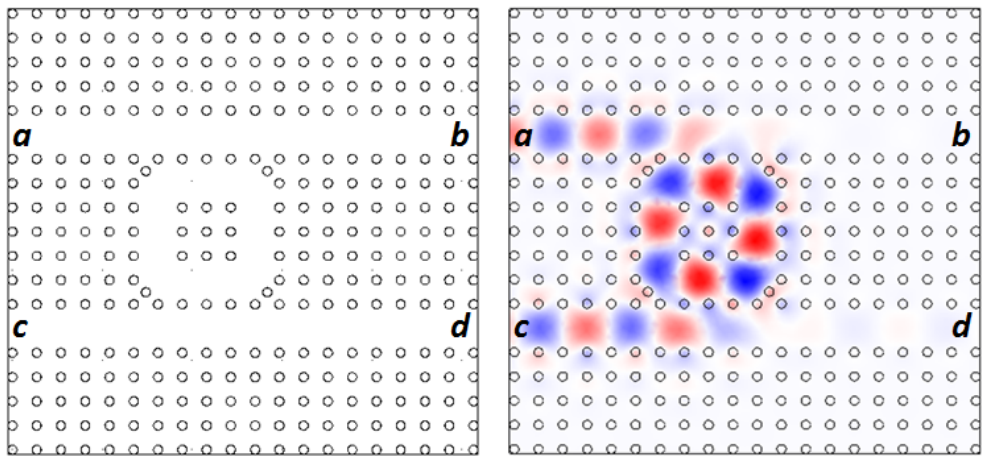

Fig. 4. Superconducting photonic ring resonator based on manipulation of periodic structure: (a) Schematic showing the ring resonator structure (b) Applying optical energy to the port A and receiving at port D with high spectral selectivity

Another method to implement photonic devices based on photonic crystals could be tuning the lattice constant in some regions of photonic device.

A high quality factor photonic nanocavity would be an appropriate candidate for the instance of implementing this category of optical devices which strongly confines optical energy in an absolutely compact dimension.

The cavity structure is sketched in Fig. 5, in which photonic crystal III, having an averaged lattice constant $a'_2 = (a_1 + a_2)/2$ is inserted between photonic crystals I and II. [16]

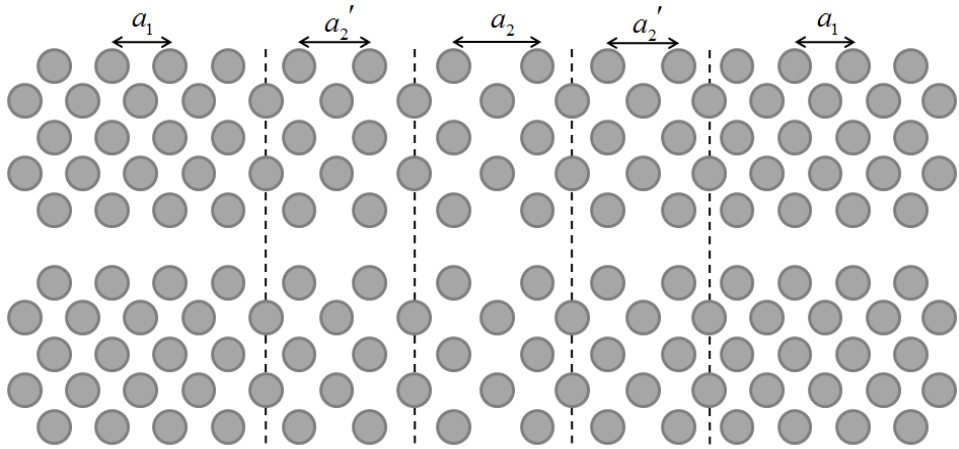

Fig. 5. Schematic of a nanocavity with multistep photonic heterostructure.

Fig. 6 presents the electric field profile of the multistep photonic heterostructure cavity along the waveguide direction for the case of full superconductivity suppressed rods. The considered lattice constants in the simulation are as follows: $a_1$=410 nm, $a_2$=420 nm. The cavity $Q$-factor achieved for this superconducting photonic device is $1.2\times10^7$, which it is in the order of that of non-superconducting nanocavity.

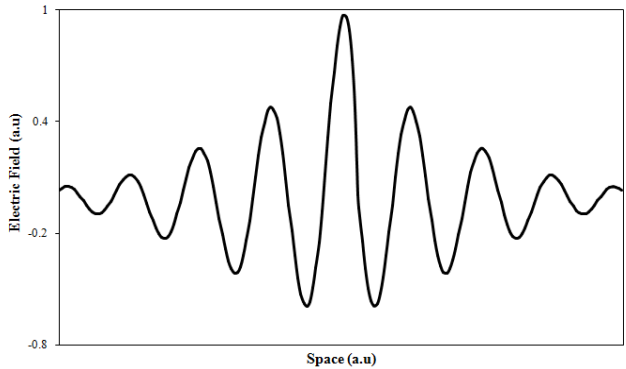

Fig. 6. Electric field distribution of heterostructure cavity and its profile along the waveguide direction

Other type of photonic devices can be implemented by changing the radius of some rods. The edge-effect is another issue which can be studied by this method. In this regards, the effects of the radius variation at the zigzag edges of a honeycomb photonic crystal on the light propagation is selected. [17]

Fig. 7 shows the honeycomb lattice of the totally suppressed rods in the homogeneous superconducting media. The rods at the edge of the device have radius of $r_3$=0.34$d$, and the radius of the inner rods are $r_1$=0.35$d$ and $r_2$=0.4$d$, respectively.

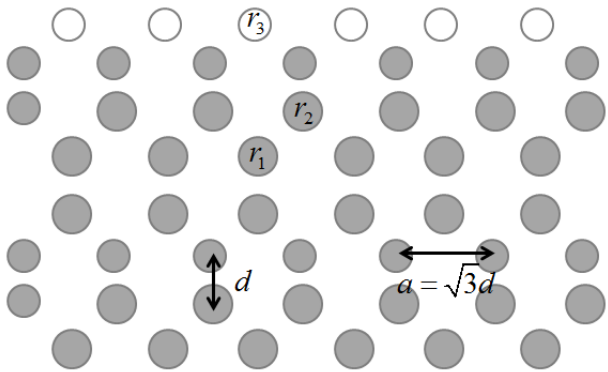

Fig. 7. Schematic of a semi-infinite honeycomb lattice of dielectric rods with modified radius of the rods at the zigzag edge.

The dispersion curve of edge states is plotted in Fig. 8 for a zigzag edge of this superconducting photonic device. The corresponding edge states have an approximately sinusoidal dispersion which leads to a low group velocity, $v_g$, with small variations in a wide frequency range.

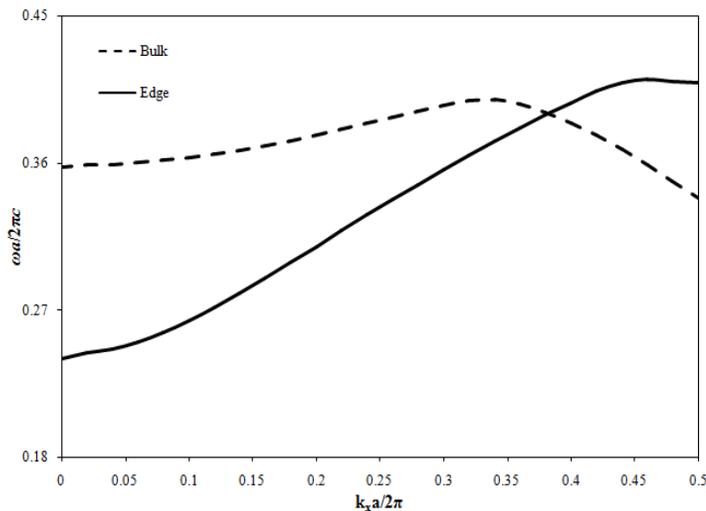

Fig. 8. Band structure of the bulk and edge states in the modified honey comb lattice

## IV. Conclusion

We theoretically demonstrated the formation of two dimensional photonic crystals composed of superconductor thin film under selective light irradiation. Due to the suppression of superconductivity, the under-emission parts could behave as the rods and provide a different permittivity. The proposed photonic crystal exhibits wide tunability by changing the light intensity and illumination pattern. We showed that the photonic band gap and the center frequency increase with enhancement of the light intensity. Also, the effect of fill factor and lattice structure on the photonic crystal properties was investigated. Moreover, modification of the illumination pattern makes it possible to develop different types of photonic devices. Manipulation of periodic structure, variation of the lattice constants and changing the radius of rods were proposed to implement ring resonator, high Q-factor nanocavity heterostructure, and optical delay line, respectively. The results reveal that the photonic characteristics of optical devices are strongly dependent on the irradiation properties, which is ideal to fabricate ultra-fast optically-controlled photonic devices.